\documentclass[aps,prl,twocolumn,groupedaddress,floatfix]{revtex4}

\usepackage{graphicx}
\usepackage{subfigure}
\usepackage{epsfig}
\usepackage{amsmath}
\usepackage{bbm}

\def\be{\begin{equation}}
\def\ee{\end{equation}}
\def\bea{\begin{eqnarray}}
\def\eea{\end{eqnarray}}

\begin{document}


\title{Electronic correlations in graphite and carbon nanotubes from Auger spectroscopy}


\author{E. Perfetto$^{1}$, M. Cini$^{2,3}$, S. Ugenti$^{2,3}$, P. Castrucci$^{1,2}$, M. Scarselli$^{1,2}$ and  M. De Crescenzi$^{1,2}$}
\affiliation{ $^{1}$  Unit\`{a} CNISM, Universit\`{a} di Roma Tor
Vergata, Via della Ricerca Scientifica 1, I-00133 Rome, Italy.
\\$^{2}$Dipartimento di Fisica, Universit\`{a} di Roma Tor
Vergata, Via della Ricerca Scientifica 1, I-00133 Rome, Italy.\\
$^{3}$Istituto Nazionale di Fisica Nucleare - Laboratori Nazionali
di Frascati, Via E. Fermi 40, 00044 Frascati, Italy. }

\author{F. Rosei and M. A. El Khakani}
\affiliation{ Institut National de la Recherche Scientifique,
INRS-\'{E}nergie, Mat\'{e}riaux et T\'{e}l\'{e}communications, Varennes, Quebec J3X 1S2, Canada }


\date{\today}

\begin{abstract}
We have determined the screened on-site Coulomb repulsion in
graphite and single wall carbon nanotubes by measuring their Auger
spectra and performing a new theoretical analysis based on an
extended Cini-Sawatzky approach where only one fit parameter is
employed. The experimental lineshape is very well reproduced by
the theory and this allows to determine the value of the screened
on-site repulsion between $2p$ states, which is found to be $2.1$
eV in graphite and $4.6$ eV in nanotubes. The latter is robust by
varying the nanotube radius from 1 to 2 nm.

\end{abstract}

\maketitle

Carbon nanostructures continue to be an intense field of both
fundamental and applied research because of the recent discoveries
of several of their unusual physical properties. Among these one
can recall \textit{(i)} the observation of the anomalous integer
quantum Hall effect in planar graphene\cite{novo,zhang}
\textit{(ii)} the measurement of superconductivity at 11.5 K in Ca
intercalated graphite and \textit{(iii)} intrinsic
superconductivity in multi-wall\cite{tk} and
ultra-small\cite{tang} carbon nanotubes at temperatures of 12 and
15 K respectively. In the light of these unprecedented properties
and related new physics, the study and the quantitative estimate
of electronic correlations in these carbon nanostructures are of
paramount fundamental importance. In fact, in one-dimensional
conductors, like metallic nanotubes, the electronic interactions
have a dramatic impact on their electronic properties, giving rise
to the so-called Luttinger liquid behavior. This manifests in the
power-law dependence of observables such as the tunneling density
of states (DOS), of which suppression at low energies has been
observed in conductance measurements\cite{yao,exp}. More
importantly the accurate estimate of the screened Coulomb
repulsion is a challenging problem that should be dealt within any
theoretical study aiming at addressing the question of
superconductivity.

Auger electron spectroscopy is a powerful experimental tool which
permits the characterization of the effective interaction between
electrons in solids. In particular the Auger lineshape is
proportional to the 2-particle interacting DOS as a consequence of
two valence holes creation on the same lattice site caused by the
X-ray photoemission of a deep core electron. Several attempts have
been made to interpret the Auger spectra of amorphous graphite
\cite{houston} and highly oriented pyrolitic graphite
(HOPG)\cite{Dementjev} but a satisfactory description is still to
come. Moreover only few experimental data on single wall carbon
nanotubes (SWCNTs) Auger lineshape are available \cite{Dementjev}.
Furthermore, no theoretical effort introducing Coulomb repulsion
in SWCNTs has been attempted so far.

In this paper we report on the comparative study of  the Auger
spectra of HOPG and SWCNTs. Through a new theoretical analysis of
the Auger experimental data, we provide an accurate estimate of
the on-site screened repulsion in both carbon structures. The
access to this quantity is key in realistic local density
approximation (LDA) $+U$ calculations and in any low-energy
interacting theory of the honeycomb lattice where only $\pi$
Dirac-like electrons are considered.

SWNCTs were synthesized by ablating a CoNi-doped graphite target,
using a pulsed Nd:YAG laser in the superposed double pulse
configuration \cite{Ali}. Raman spectroscopy indicated that the
tube are single wall, characterized by a low degree of defects and
with diameters in the range 1.2-1.3 nm. \cite{Ali}. This is
consistent with transmission electron microscopy (TEM)
observations that, though showing tubes aggregated in bundles of
various dimension and twisting, allowed us to measure a tube
diameter of 1.2$\pm$ 0.1nm through a statistical analysis
\cite{MP1}. Moreover electron energy loss spectroscopy performed,
by using the TEM apparatus, directly on SWCNTs bundles at the Co
and Ni L$_2,3$ edges did not detecte any traces of these
catalysts. \cite{MP2} A droplet of the synthesis product was
diluted in isopropyl alcohol and dispersed on a metallic surface.
A freshly cleaved HOPG sample was used for measuring the
core-valence-valence (KVV) Auger features. The Auger spectra were
acquired using an Al K$\alpha$ (1486.6 eV) monochromatic x-ray
source with a resolution of about 1 eV. The obtained experimental
spectra are shown in Fig.1 after subtraction of secondary electron
background.

The Auger lineshape of solids can be calculated by using the
so-called 2-step approach, in which the photoemission and the
Auger decay are considered as independent events. In absence of
significant electronic correlations, the computation of KVV Auger
spectrum reduces to the self-convolution of the 1-particle valence
DOS. If moderate or strong (compared to the bandwidth) on-site
repulsion is present, the lineshape can be calculated by means of
the Cini\cite{cini1}-Sawatzky\cite{saw} approach.


Following Ref.\cite{cini1}, the Auger current $J$ reads
\begin{equation}
J=\sum_{\alpha_1,\alpha_2,\alpha_3,\alpha_4,\sigma}
A_{\alpha_1,\alpha_2,\alpha_3,\alpha_4,\sigma}
D_{\alpha_1,\alpha_2,\alpha_3,\alpha_4,\sigma}(\omega)
\label{curraug}
\end{equation}
where $\alpha_{i}$ denote all the single-particle valence orbitals
available in the solid, $A$ is the so-called Auger matrix element
given by
\begin{eqnarray}
A_{\alpha_{1},\alpha_{2},\alpha_{3},\alpha_{4},\sigma}&=& \sum_{k}
\langle  v | d_{\alpha_{1} \uparrow} d_{\alpha_{2} \sigma}|
\frac{e^{2}}{r}| d^{\dagger}_{c \sigma _{c}}
d^{\dagger}_{k \sigma_{k}} | v \rangle \nonumber \\
& \times & \langle v | d_{c \sigma _{c}} d_{k \sigma_{k}} |
\frac{e^{2}}{r}|d^{\dagger}_{\alpha_{3} \uparrow}
d^{\dagger}_{\alpha_{4} \sigma}| v \rangle
 \end{eqnarray}
 with $k,c$ and $\sigma_{k}, \sigma_{c}$ denoting the Auger electron and core orbitals
and spin respectively. $D$ is the 2-particle interacting DOS
\begin{equation}
D_{\alpha_1,\alpha_2,\alpha_3,\alpha_4,\sigma}(\omega)= \langle v
| d_{\alpha_{1} \uparrow} d_{\alpha_{2} \sigma}| \delta(\omega-H)|
d^{\dagger}_{\alpha_{3} \uparrow} d^{\dagger}_{\alpha_{4} \sigma}
| v \rangle \, . \label{intdos}
 \end{equation}
 where $H$ is the interacting hamiltonian of the solid.
 Here we denote by $|v\rangle$ the hole-vacuum and by $d^{(\dagger)}_{i}$
 the annihilation (creation) operator of a hole in spin-orbital $i$.
$D$ is obtained as usual from the anti-hermitian part of the
2-particle Green's function
$G_{\alpha_1,\alpha_2,\alpha_3,\alpha_4,\sigma}(\omega)$ which
obeys the matrix Dyson\cite{cinimatrix} equation
\begin{equation}
G_{\sigma}=G_{\sigma}^{(0)}[1+ U_{\sigma} G_{\sigma}^{(0)}]^{-1}
\, , \label{green}
\end{equation}
where $G^{(0)}$ is the noninteracting 2-hole Green's function and
$U$ is the matrix of screened on-site repulsion for valence
states. The screened interaction differs from the bare atomic one,
defined as
\begin{equation}
U^{b}_{\alpha_1,\alpha_2,\alpha_3,\alpha_4,\sigma}= \langle v |
d_{\alpha_{1} \uparrow} d_{\alpha_{2} \sigma}| \frac{e^{2}}{r}|
d^{\dagger}_{\alpha_{3} \uparrow} d^{\dagger}_{\alpha_{4} \sigma}
| v \rangle \, .
 \end{equation}
The evaluation of $U$ starting from the atomic value $U^{b}$ is
generally a delicate task. In the following we discuss the
phenomenological approach we have adopted to  determine this
quantity. The Cini-Sawatzky approach works quite well in closed
(or almost closed)-band systems like zinc and copper, where the
ladder approximation leading to Eq. (\ref{green}) provides an
exact result. However, if the Fermi level crosses the middle of
the conducting band, the computation of the Auger current becomes
a remarkably challenging many-body problem, which usually cannot
be solved by evaluating Green's functions\cite{gunnsch}.

\begin{figure}
\begin{center}
\mbox{\epsfxsize 6.0cm \epsfbox{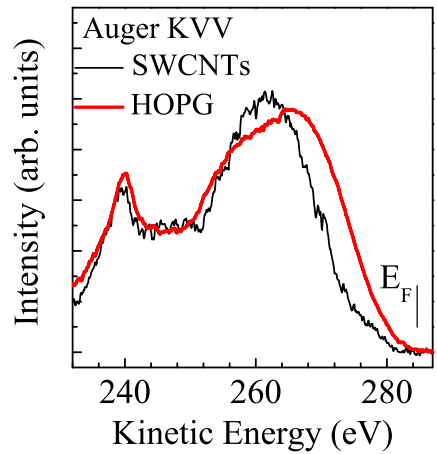}}
\end{center}
\caption{(Color online) Experimental KVV Auger spectra of HOPG
graphite (bold red curve) and SWCNTs with average diameter of
$1.3$\textit{n}m (black curve)} \label{spectrummexp}
\end{figure}

In the light of this, the theoretical study of Auger spectra of
HOPG and SWCNTs is indeed far from straightforward because the
$\sigma$ and $\pi$ bands are half filled. However in these systems
some special features (which are discussed hereafter) allow the
use of closed-band theory, with slight but crucial modifications.

First we observe that the DOS is largely suppressed in the
proximity of the Fermi level, so that screening is not very
efficient. This implies a static renormalization of the bare
interaction $U^{b}$ which must be used in the theory. Second, we
recall that the bonding portion of the $\sigma_{s,p}$ bands is
separated by several eV from the antibonding part located above
the Fermi level. As long as such a separation is larger than the
effective interaction, one can treat the band as if it was closed,
thus justifying the approach reported by Cini\cite{cini1}, where
no structural modification  is needed for the interacting Green's
function in Eq. (\ref{green}). However, the situation is different
for the $\pi$ band, where the bonding and antibonding portions are
separated by a very small region with a small DOS. Here Cini's
approach can not be used without appropriate modifications. In
this case the contribution to the Auger spectrum originating from
$\pi$ and mixed $\pi - \sigma$ holes would be strongly influenced
by open-band effects. It is also expected that such a region
should reveal the principal differences between the spectra of
HOPG and SWCNTs.  In fact screening and excitonic
effects\cite{houston2} and Luttinger liquid properties in SWCNTs
are expected to lead to a quite different behavior of electrons in
proximity of the Fermi level due to the different dimensionality.
This conjecture seems to be confirmed by the experimental data.
Indeed, the $\pi$ and mixed $\pi - \sigma$ portion of spectrum
(i.e. $\omega \gtrsim 250$ eV) show clear differences between HOPG
and SWCNTs, while in the $\sigma_{s}$ region (i.e. $\omega
\lesssim 250$ eV) the two spectra are quite similar. In particular
for $ 250$ eV $ \lesssim \omega \lesssim 280$ eV the lineshape of
SWCNTs is narrower with vanishing and much weaker intensity in
proximity of the Fermi level, as compared to the one for graphite.
This fits well with a scenario where the screening properties of
$\pi$ electrons are less efficient in SWCNTs.

Within the closed-band theory, the Auger spectrum is obtained by
taking the Auger matrix elements  and the on-site interactions
from atomic calculations, which neglect solid state effects. On
this basis, one introduces the static screening operated by the
closed-band system simply by rescaling all the $F^{(0)}(i,j)$
Slater integrals that enter the bare $U^{b}$, such that
$F^{(0)}(i,j) \rightarrow F^{(0)}(i,j)-W$. $W$ can be taken as the
unique free fitting parameter of the theory. Alternatively $W$ can
be also estimated within the Random Phase Approximation or
\textit{ab initio} methods\cite{abinitioU}. The only ingredient
which accounts that the Auger holes are in the solid is the
noninteracting 1-particle DOS $\rho^{(0)}(\omega)$. Its
self-convolution $D^{(0)}(\omega)= \int d \varepsilon
\rho^{(0)}(\varepsilon) \rho^{(0)}(\omega-\varepsilon)$ and the
corresponding Hilbert transform  build the noninteracting $G^{0}$
entering Eq. (\ref{green}).

Cini's approach should in principle be completed by introducing
the effect of off-site interaction. Experiments on Au
\cite{offsite1} showed that there is a shift of 1.2 eV between the
profile predicted by the above theory and experiment. The shift is
2.4 eV in the case of  Ag \cite{offsite2}. This was explained in
terms of the off-site interaction. In the two-hole resonance there
is an important amplitude that the holes sit on neighboring sites,
and including the nearest-neighbor interaction into the theory
yields an almost rigid shift close to the experimental
one\cite{offsite3}.

\begin{figure}
\begin{center}
\mbox{\epsfxsize 8.0cm \epsfbox{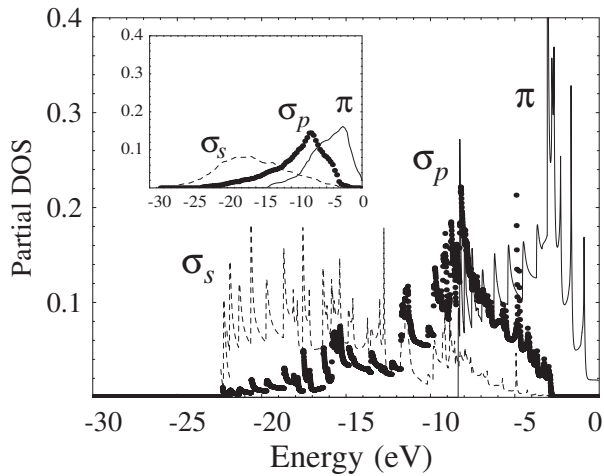}}
\end{center}
\caption{1-particle partial DOS of (10,10) SWCNT (diameter close
to 1.3 nm) obtained by the tight binding method of
Ref.\onlinecite{dress}. The inset shows the same quantity for
graphite, taken from Ref.\onlinecite{houston}. The Fermi level
corresponds to zero-energy and the antibonding part is not
displayed. } \label{1pdos}
\end{figure}

In the following we will phenomenologically consider the open-band
effects by introducing orbital-dependent \textit{form factors}
$f_{\alpha_1,\alpha_2,\alpha_3,\alpha_4}$.  This must be
introduced to correct all the quantities measuring \textit{local}
properties expressed by $\langle v | d_{\alpha_{1}}
d_{\alpha_{2}}| O | d^{\dagger}_{\alpha_{3} }
d^{\dagger}_{\alpha_{4}} | v \rangle $ where $O$ is a local
observable.  Therefore the effective on-site repulsions
$U_{\alpha_1,\alpha_2,\alpha_3,\alpha_4,\sigma} $  (where $F^{0}$
has been already rescaled by $W$) and the matrix elements
$A_{\alpha_1,\alpha_2,\alpha_3,\alpha_4,\sigma}$ will be corrected
by a common multiplying factor
$f_{\alpha_1,\alpha_2,\alpha_3,\alpha_4}$. In our case the
$\alpha_{i}$ states are $\sigma_{s}, \sigma_{x}, \sigma_{y}, \pi$.
The form factor $f$ takes into account that the $2s$ states of
carbon behave as if they were atomic, while the $2 p$ ones are
delocalized in the lattice. The latter can use only $1/2$ of the
total $\sigma_{p}$ and $\pi$ states to form occupied localized
states because the $p$-bands are half-filled. Therefore we have
three independent $f$ factors corresponding to having \textit{(i)}
four $\sigma_{s}$ orbitals , \textit{(ii)} two $\sigma_{s}$ and
two $\sigma_{x,y},\pi$ orbitals, \textit{(iii)} four
$\sigma_{x,y},\pi$ orbitals in the quartet $\{
\alpha_1,\alpha_2,\alpha_3,\alpha_4 \}$. According to the above
discussion the three independent form factors are estimated to be
$f_{ssss} \approx 1$, $f_{sspp} \approx 1/2$ and $f_{pppp} \approx
1/4$. We will show that this choice works quite well in the case
of HOPG, while we need $f_{pppp} \approx 1/2$ to reproduce the
Auger spectrum of nanotubes.  Indeed in nanotubes the geometry
constrains the holes and this could be the reason for a larger
$f_{pppp} $ than in graphite. It is worthwhile to note that the
analysis of Ref.\cite{offsite3} does not apply to $p$ holes and in
fact no shift is seen in this case (the pairs presumably extend
further than a nearest neighbor distance). A shift could be
present in the $KL_{1}L_{1}$ case, but we cannot tell since there
is a single peak there.


We proceed by evaluating the noninteracting 1-particle DOS
$\rho^{(0)}$ for each kind of valence state. In the case of HOPG,
we use the DOS from Ref.\onlinecite{houston} which is taken from
experiments. For SWCNTs we performed a tight binding
calculation\cite{dress} including both $2s$ and $2p$ orbitals, but
neglecting overlap integrals for simplicity. The result for a
typical (10,10) armchair nanotube with diameter close to 1.3 nm is
shown in Fig.\ref{1pdos} together with the DOS of HOPG.
For the Auger matrix elements, we used the (spin-independent)
values $A_{ssss}=0.8,A_{sspp}=0.5,A_{pppp}=1.0$ which are obtained
by atomic calculations\cite{houston} and hence apply to both
graphite and carbon nanotubes. The bare (atomic) on-site Coulomb
repulsions are obtained by appropriate combinations of the Slater
integrals $F^{(0,2)}(i,j)$ and $G^{(1)}(i,j)$\cite{weiss} found in
the literature\cite{mann}. The independent bare interactions are
(in eV) $U^{b}_{ssss \downarrow}=15.5$, $U^{b}_{sxsx
\downarrow}=15.0$, $U^{b}_{sxxs \downarrow}=1.5$, $U^{b}_{\pi \pi
\pi \pi \downarrow} (\equiv U^{b}_{pppp}) =14.6$, $U^{b}_{xx \pi
\pi \downarrow}=-0.1$, $U^{b}_{x \pi x \pi \downarrow}=13.9$,
$U^{b}_{x \pi \pi x \downarrow}=0.8$, $U^{b}_{ssxx
\downarrow}=11.9$, $U^{b}_{sxsx \uparrow}=$, $U^{b}_{x \pi x \pi
\uparrow}=13.1$. As discussed above, these values must be
corrected by solid state effects. This is done by subtracting the
screening constant $W$ from the $F^{(0)}(i,j)$ Slater integrals
and multiplying the resulting $U$ and $A$ matrix elements by the
$f$ factors ($W$ being the the only fitting parameter of our
approach).

\begin{figure}
\begin{center}
\mbox{\epsfxsize 6.0cm \epsfbox{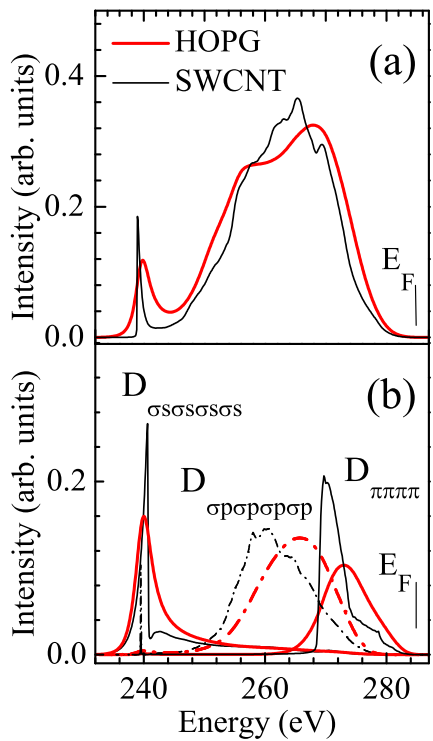}}
\end{center}
\caption{(Color online) (a) Theoretical lineshape [computed from
Eq. (\ref{curraug})] of KVV Auger spectrum for HOPG (bold red) and
for SWCNTs (black) curve; (b) Diagonal contributions of the
interacting DOS for HOPG (bold red) and for SWCNTs (black) where
the two valence holes have the same symmetry. The
$D_{\sigma_{p}\sigma_{p}\sigma_{p}\sigma_{p}}$ contribution is
understood as the sum
$D_{\sigma_{x}\sigma_{x}\sigma_{x}\sigma_{x}}+D_{\sigma_{y}\sigma_{y}\sigma_{y}\sigma_{y}}$.
The $x$-axis displays kinetic energy, obtained by shifting the
position of the Fermi level in Fig.2 of 284.6 eV, which is the
binding energy of $1s$ core hole.} \label{spectrumt}
\end{figure}

The theoretical spectra of HOPG and SWNCTs were computed by
solving a $16 \times 16$ matrix problem for $\sigma = \downarrow$
and a $6 \times 6$ problem for $\sigma = \uparrow$, as shown  in
Eq. (\ref{green})\cite{cinimatrix}. The final result is plotted in
Fig.\ref{spectrumt}a, where the best fittings yielded the
respective values $W_{\mathrm{HOPG}} = 6.0$ eV and
$W_{\mathrm{SWCNT}} = 5.5$ eV for HOPG and SWCNTs. The agreement
between theory and experiment is quite good, and is particularly
satisfactory for graphite.

These values permit the determination of the most relevant
parameter of our model, which is the screened on-site repulsion
between the $2p$ states. Thus, the  best fitting for $W$ yields
$U_{pppp} = 2.1$ eV for HOPG and $U_{pppp} = 4.6$ eV for SWCNTs.
This result gives rise to the lack of features close to the Fermi
level for SWCNTs, making the Auger spectrum more symmetric and
narrower than that of HOPG. This is understood by looking at
Fig.\ref{spectrumt}b, which  shows the diagonal contributions of
the interacting DOS according to Eq. (\ref{intdos}), where the
valence holes were taken in the same state. The off-diagonal
contributions are not shown for the sake of clarity but are
essential to reproduce the experimental spectra.

Concerning the lineshapes, the most striking feature is the narrow
structure at 240 eV, which also appears as a shoulder in the
spectrum reported by Houston et al.\cite{houston}. This peak was
assigned to a plasmon replica of the main structure at 265 eV
produced by a plasmon with an energy $\omega_{p}=27$ eV.
Conversely we interpret the narrow structure as a quasi-two-hole
resonance produced by two $\sigma_{s}$ Auger holes. This is
consistent with the predicted values of the screened on-site
repulsion between $\sigma_{s}$ holes, which are $U_{ssss} = 9.5$
eV and $10.0$ eV for HOPG and SWCNTs respectively. The
noninteracting $D^{(0)}_{ssss}$ has a maximum at
$\varepsilon_{ss}=252$ eV (graphite) and $251$ eV (nanotube) and
therefore a narrow structure around $\varepsilon_{ss}-U_{ssss}
\approx 241$ eV in the interacting $D_{ssss}$ is correctly
expected. Since the $\sigma_{s}$-bandwidth is of $\sim 20$ eV a
full splitoff two-hole resonance cannot happen, but a strongly
distorted band-like behavior occurs (see Fig. 1b of
Ref.\onlinecite{cini1}). It is worth noting that Auger spectrum
from a sample consisting of SWCNTs with average diameter of 2nm
does not show significant changes with respect to that reported in
Fig.\ref{spectrummexp}. Moreover, by performing a similar
theoretical analysis on a (20,20) SWCNT no substantial changes can
be found for the values of the correlation interaction. This means
that the values we obtain for the correlation in SWCNTs have a
very small dependence on the nanotubes diameter.

In conclusion the lineshape of the Auger spectra for HOPG and
SWCNTs have been interpreted in terms of a new theoretical
approach using a single fitting parameter. The $U_{pppp}$ Coulomb
repulsion results doubled passing from HOPG to SWCNTs. This
explains the sizeable shift of the Auger feature at high kinetic
energy measured for SWNTs, as compared to HOPG. Finally we point
out that the increase of the $U_{pppp}$ value is consistent with
the theoretical prediction\cite{ciniprb2007} of the enhancement of
the superconductive critical temperature observed recently in
carbon nanotubes.

E.P. was supported by CNISM. M.C. and S.U. acknowledge support by
the Italian Ministry  Murst  under the PRIN  code:
$2005021433$$\_$$002$ year: 2005. M.A.E., P.C., M.S., M.D. thank
the Italian Foreign Affairs Ministry through Promotion and
Cultural Cooperation Management for financial support. F.R. is
grateful to FQRNT (Qu\'{e}bec) and the Canada Research Chairs
program for partial salary support.

\end{document}